\documentclass[aaspp4,epsf]{article}
\usepackage{osa2}

\begin{document}
\title{A New Pupil for Detecting Extrasolar Planets}

\author{D. N. Spergel}
\address{Department of Astrophysical Science, Princeton University, Princeton, NJ
08544}
\address{Institute for Advanced Study, Princeton, NJ 08540}
\email{dns@astro.princeton.edu}

\begin{abstract}The challenge for optical detection of terrestial planet is the 25 magnitude
brightness contrast between the planet and its host star. This paper introduces a
new pupil design that produces a very dark null along its symmetry axis. 
By changing the shape of the pupil, we can control the depth and location of
this null. This null can be further enhanced by combining this pupil with a
rooftop nuller or cateye nuller and an aperture stop. The performance of the
optical system will be limited by imperfections in the mirror surface.  If
the star is imaged with and without the nuller, then we can characterize
these imperfections and then correct them with a deformable mirror.  The
full optical system when deployed on a 6 $\times$ 10 m space telescope is capable
of detecting Earthlike planets around stars within 20 parsecs.  For an
Earthlike planet around a nearby stars (10 parsecs), the telescope can
characterize its atmosphere by measuring spectral lines in the 0.3 - 1.3
micron range.
\end{abstract}

\centerline{Submitted to Applied Optics}

\section{Introduction}

The detection and characterization of Earthlike planets is one of a NASA's
highest priorities for the coming years. Currently, NASA's plans focus on
building large mid-infrared nulling interferometers that aim to detect
thermal emission from planets and characterize the absorbtion lines in their
atmosphere\cite{TPF}. There is an alternative approach: a
large optical telescope with a suitable coronagraph may be capable of
detecting reflected light from the planets\cite{Ford}.  Since this
large telescope would have a high quality optical surface, it would also be
capable of 10 milliarcsecond imaging and would be a powerful instrument for
near UV spectroscopy.  There are a wide range of astronomical problems that could be addressed with
a large optical/near UV space telescope\cite{Ford,Shull}. This large space telescope would
be a boon to astrophysics!

Brown and Burrows\cite{Brown} discuss searching for extrasolar planets with the
Hubble Space Telescope.  They emphasize the challenge for extrasolar planet
searches: detecting the planet's weak signal against the star's bright
light.  They show that there were two dominant sources of noise in the
observation: diffracted light due to the finite size of the aperature and
scattered light due to imperfections in the mirror surface.

This paper is an attempt to design a system that is optimized for extrasolar
planet detection.  In section 2, I introduce a new pupil design and show
that it can significantly reduce diffracted light along one axis of the
system and discuss optimizing the pupil shape for terrestial planet finding.
The optical system can be enhanced by combining this pupil with a
coronagraph.  In section 3,
I discuss the scattered light problem and show how we can use a
spectrograph placed along the symmetry axis of the system to characterize
the mirror imperfections.  A deformable mirror can correct these
imperfections and reduce the stellar scattered light at the position of the
planet.  Section 4 discusses using this system to detect Earthlike planets
around nearby stars and concludes the paper.

\section{A New Pupil Design}

\subsection{Analytical Pupil Shape}

\begin{figure}[ht] 
\centerline{\scalebox{.4}{\includegraphics{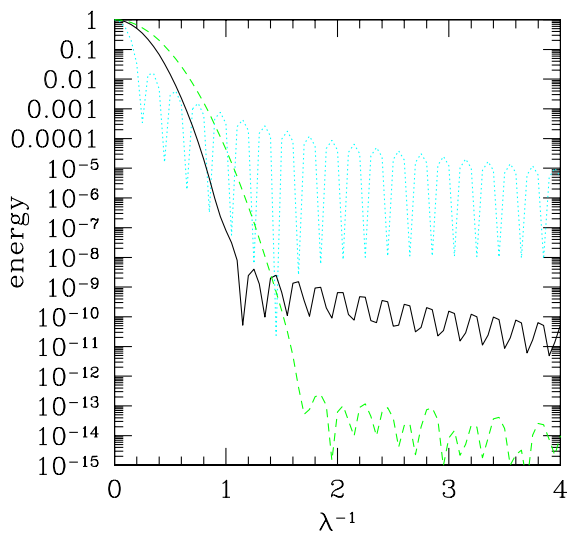}}}
\caption{Pupil diffraction pattern.  The y axis shows the 
incident flux normalized to the central value for different pupil shapes.  The x axis is measured
in inverse wavenumbers (microns$^{-1}$) for a pixel seperated
0.1 arcseconds from the point source along the $\xi$ axis.  
The blue dotted line is for a circular aperature.  The solid line
is for a shaped aperature (equation 2) with $\alpha = 2.7$.  The green
dashed line is for a shaped aperature with $\alpha = 3.0$.}
\end{figure}

\begin{figure} [ht]
\centerline{\scalebox{.4}{\includegraphics{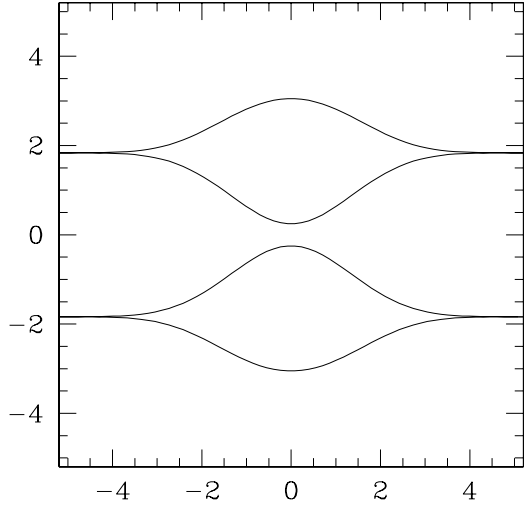}}}
\caption{Pupil Shape.  This figure shows the two piece version
of the pupil design. The axes  are in meters}
\end{figure}

\begin{figure}[ht]
\centerline{\scalebox{.4}{\includegraphics{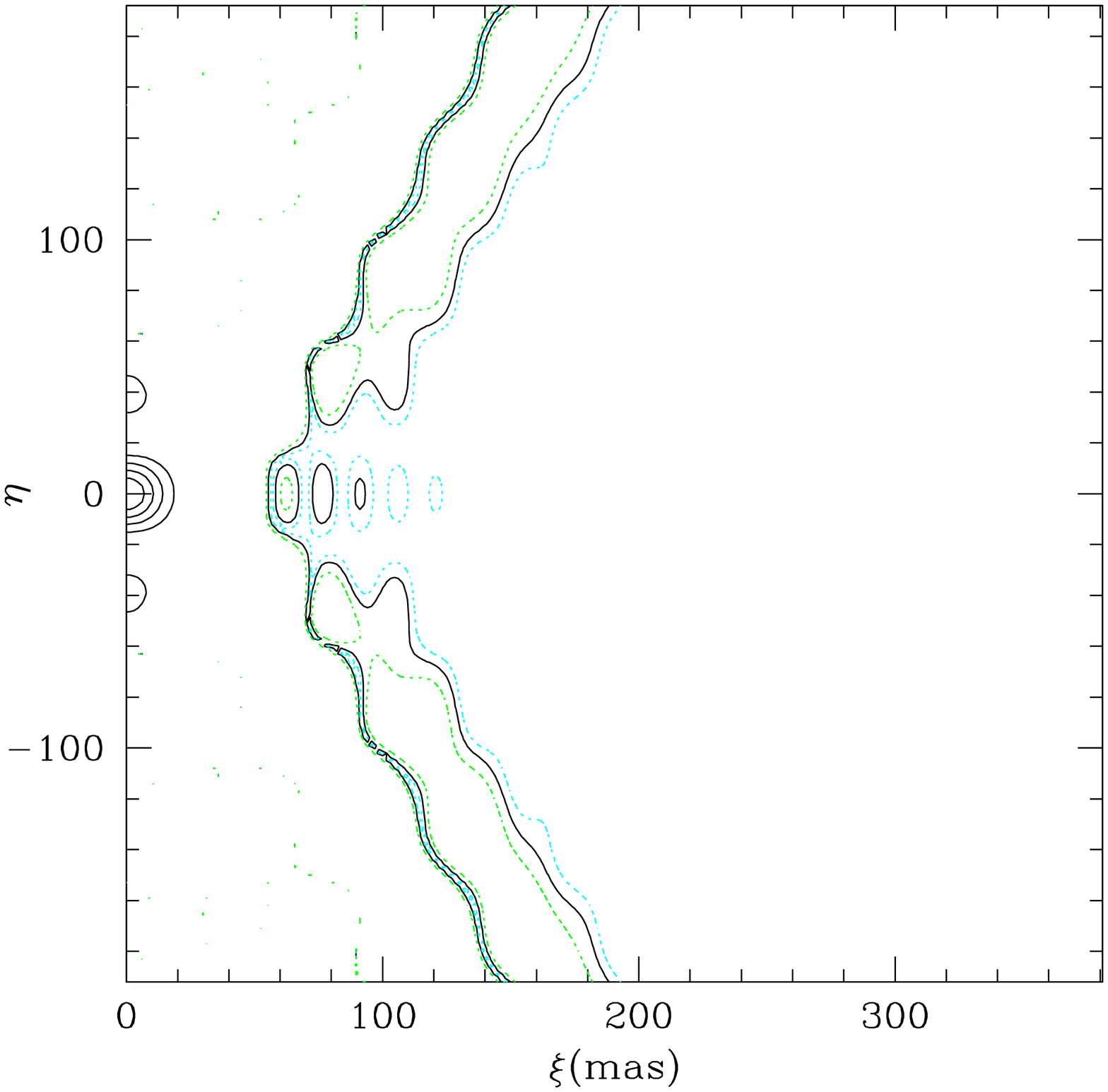}}}
\caption{Energy distribution.  Near the peak, the solid contours show
relative intensity levels of 0.2, 0.4, 0.6 and 0.8 times the peak intensity.
Along the x-axis, the dotted, solid black, and dotted contours
show regions with relative intensities of $10^{-9}$, $3\times 10^{-10}$
and $10^{-10}$}
\end{figure}

In this section, I introduce a pupil designed to minimize the diffracted
light from the host star along its symmetry axis.

In the scalar optics limit, the diffraction pattern produced  point source
observed through a pupil  at focus is the Fourier transform of its shape\cite{BW}: 
\begin{equation}
I(\xi ,\eta )=\int_{A(x,y)}dxdy\exp \left[ ik\left( x\xi +y\eta \right) %
\right]
\end{equation}
Throughout this paper, $k$ is the photon wavenumber, $\xi $ and $\eta $ are
positions in the focal plane, and $x$ and $y$ are positions along the mirror
surface.  Along the $\xi $ axis, the optical response depends on the width
of the pupil: 
\begin{equation}
I(\xi ,\eta )=\int_{{}}dx\exp \left( ikx\xi \right) \left[ y_{width}(x)-%
\frac{k^{2}\eta ^{2}}{2}\int_{A(x,y)}y^{2}dy+\frac{k^{4}\eta ^{4}}{24}%
\int_{A(x,y)}y^{4}dy+...\right]
\end{equation}
where the odd terms in the expansion vanish as long as the pupil is
symmetric about the x axis.  For a circular aperature of radius, $R,$ $%
y_{width}(x)\;=\sqrt{R^{2}-x^{2}}$ and the diffraction pattern is the
familiar Airy pattern. For a circular pupil, the energy recieved falls
off as $\theta ^{-3}$.

If the width of the pupil is close to a Gaussian, 
\begin{equation}
y_{width}(x)=\exp \left[ -(\alpha x/R)^{2}\right] -\exp (-\alpha ^{2})
\end{equation}
with $-R<x<R,$ then its diffraction pattern along the $\xi $ axis falls off
much faster than that of a circular pupil: 
\begin{eqnarray}
I(\xi ,0) &=&\int_{-R}^{R}dx\exp \left( ikx\xi \right) \left\{ \exp \left[
-(\alpha x/R)^{2}\right] -\exp (-\alpha ^{2})\right\} \\
&=& {\pi \over 2} {R \exp\left({-k^2 R^2 \over 4 \alpha^2}\right) \over \alpha}
\left[{\rm erf}(\alpha +IKR/2\alpha)) - {\rm erf}(\alpha -IkR/2\alpha)\right]
\nonumber  \\
&&-\exp (-\alpha ^{2})\frac{\sin (k\xi R)}{k\xi }  \nonumber
\end{eqnarray}
Figure 1 shows the energy recieved as a function of position in the focal
plane for a circular aperature and a shaped pupil with several different
values of $\alpha .$  Note the basic trade-off in the pupil design: if 
$\alpha $ is larger, then the null is deeper; however, the dark region
starts at a larger radius.

If the pupil consists of a single opening, then equation (3) fully describes
its shape.  However, a two piece pupil is specified
by an additional function, the position of the inner edge of the pupil (see
figure 2). If the inner edge of the pupil is chosen so that 
\begin{equation}
\int_{y_{inner}(x)}^{y_{inner}(x)+y_{width}(x)}y^{2}dy=\beta y_{width}(x),
\end{equation}
then the null that is both deep and wide: 
\begin{equation}
I(\xi ,\eta )=I(\xi ,0)(1-\beta k^{2}\eta ^{2}+....)
\end{equation}
Figure 2 shows the optical response of the two piece pupil and figure 3
shows an image of the two piece pupil.

The two piece pupil is well suited for a space telescope.  The secondary
mirror of the telescope and its support structure can sit in the dark region
between the two pieces of the pupil. The primary mirror can consist of two
pieces so that it can fold up and fit into fairing of the launch vehicle. 
In this folding mirror design, the fold also sits in the dark region between
the pupil.  With proper baffling, there should be not be significant
scattered light from either the fold, the secondary or its support structure.

For a multi-mirror telescope, there are many possible higher order versions
of this pupil that consist of multiple  pieces.  As long as the integrated
width of the pupil satisfies equation (3) (or the appropriately optimized
forms discussed in 2.3), the response of the pupil will be similar.  In a
multiple mirror system, not only can the first two terms in the
expansion be supressed, but also the
higher order terms in equation (2).

\subsection{Combining the Pupil with a Coronagraph}

The performance of the optical system can be
enhanced by combining the pupil with the appropriate
coronagraph.  For example, the pupil could be
combined with the achromatic interferocoronagraph\cite{Baudoz}.
In this coronagraph, the light from the star is split by a beam splitter
into two equal intensity beams.  One of the beampath experiences an
additional reflection so that when the two beams are recombined, they are
out of phase by $\pi $ and reflect around the symmetry axis. Since the
diffraction pattern is symmetric, these two beams cancel and null out the
star. In practice, there is a small optical path difference, $\Delta ,$
between the two beampaths, so that the measured signal from the star is 
\begin{equation}
I_{meas}(\xi ,\eta )=\frac{1}{2}\left[ 1+\left( 2\pi k\Delta \right) ^{2}%
\right] I(\xi ,\eta )-\frac{1}{2}I(-\xi ,\eta )
\end{equation}
Since the planet appears on only one side of the star, it is not nulled,
instead, two images of the planet are detected.

By combining this nuller with the pupil design of section (2.1), we can
combine the advantages of both and achieve a very deep null close to the
star.  If an X-shaped stop is placed in the focal plane in front of the
nuller, then it will block most of the starlight from reaching the nuller.
Thus, mirror imperfections in the nuller will not make a noticeable
contribution to the scattered light reaching the CCD or spectrograph at the
back end of the nuller.

Next, we consider a strawman optical system and assume that
we can control the optical
path difference in the nuller to better than $\lambda /6.$ This reduces the
diffracted light from the star by 10$^{-3}$.  When combined with a pupil
designed to produce a null of depth 10$^{-7}$, the diffracted light from the
star is now suppressed so that it is less than the light from the planet. 
With this combined system, we can detect with a 10 meter telescope, an Earth
twin around star 10 parsecs away at all wavelengths shorter than 1.1
microns. In the next section, we will optimize the pupil shape and improve
the wavelength coverage by 20\%.

\subsection{Optimizing the Pupil Shape}

While the analytical pupil shape is a useful first step in designing the
pupil, its shape is not optimal. the system performance can be improved by
shaping the pupil so that it achieves a null of specified depth at all
wavelengths shorter than a critical wavelength.  I have done a two step
optimization. In the first step, the width of the pupil was varied to obtain
a deep null over as wide a wavelength range as possible along the axis of
the pupil plane. In the second step, the inner edge of the pupil was
varied to minimize the energy falling into a slit placed along the axis.

Using a simulated annealling routine, I have searched function space to find
a function for the width of the mirror that creates a null of depth $10^{-7}$
at wavelengths shorter than 1.1 microns at angular seperation of 0.08
arcseconds. When this is combined with a nuller, the star's light is
suppressed by at least 10$^{-10}.$ With this criterion, a 10 meter
diameter telescope could detect the 1 micron water vapor line in an
Earth-like planet around a Solar-type star at a distance of 10 parsecs and a
seperation of 0.8 AU.

In the first optimization routine,  I represent the mirror by its width at
64 points along a grid. At each step, the mirror shape is varied at each
position by multiplying it by a random variable. The expectation value of
this random variable is 1 and its variance is varied as a function of the
iteration. For each pupil design, the program computes a quality
parameter, 
\begin{equation}
Q=\int_{k_{\min }}^{{}}dk(I(k)-I_{\min })\;\Theta (I(k)-I_{\min })
\end{equation}
where $I_{\min }$ is a fixed threshold (10$^{-7})$ and $\;\Theta $ is the
Heavyside function. If the quality factor for the new design is better
than the current best design, the program always accepts it.  If the
quality factor is poorer than the current design, the program accepts it
with a probability $\exp [-\left( Q_{new}-Q_{old}\right) /T]$, where the
system temperature is slowly lowered through the calculation. The program
ran for $10^{7}$ steps and showed only minor improvements during the final
half of the run.

In the second step in the optimization routine, I minimized the energy
falling into a 0.03 arcsecond wide slit along the axis for wavelengths
shorter than 1.1 microns. I begin with the inner edge fixed so as to
minimize the $\eta ^{2}$ terms, where $\eta $ is th distance in the focal
plane perpendicular to the slit  (This leads to a large $\eta ^{4}$ terms.)
Using the same minimization routine, I now vary the inner edge of the
profile.

\begin{figure} [ht]
\centerline{\scalebox{.4}{\includegraphics{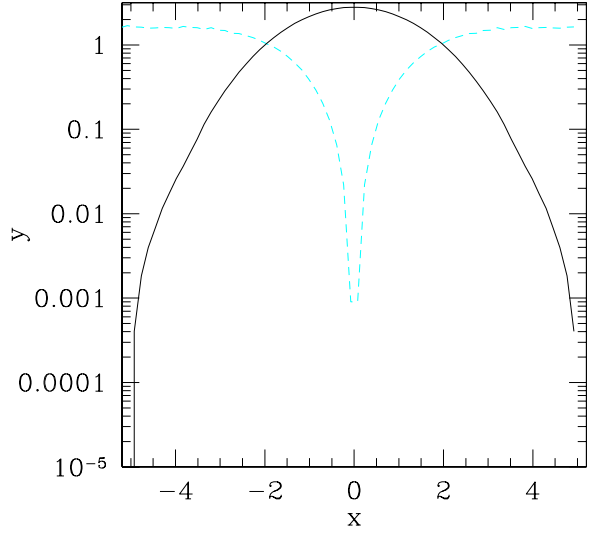}}}
\caption{Pupil Shape.  The solid line shows the width of the pupil in meters
as a function of x.  The dashed line shows the position of the inner edge.}
\end{figure}
\begin{figure} [ht]
\centerline{\scalebox{.4}{\includegraphics{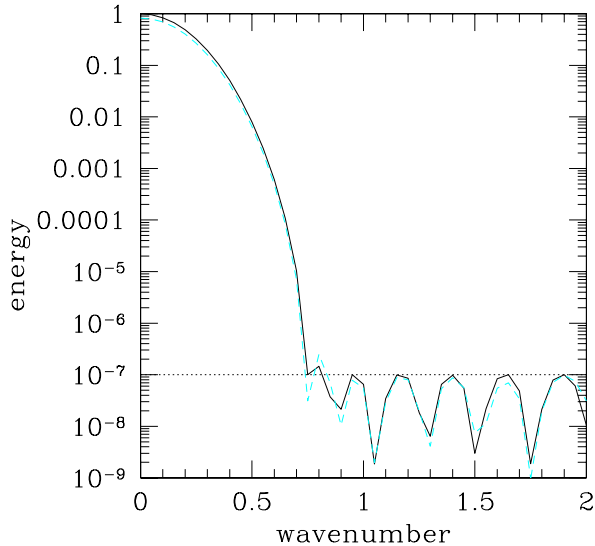}}}
\caption{The dashed line shows the amount of stellar light recieved into
a 0.08 arc second wide slit located 0.1 arcseconds from the host star.
The solid line shows the amount of stellar light recieved into a 0.005
arc second slit.  Both lines are normalized to unity at wavenumber 0.}
\end{figure}
\begin{figure}[ht]
\centerline{\scalebox{.4}{\includegraphics{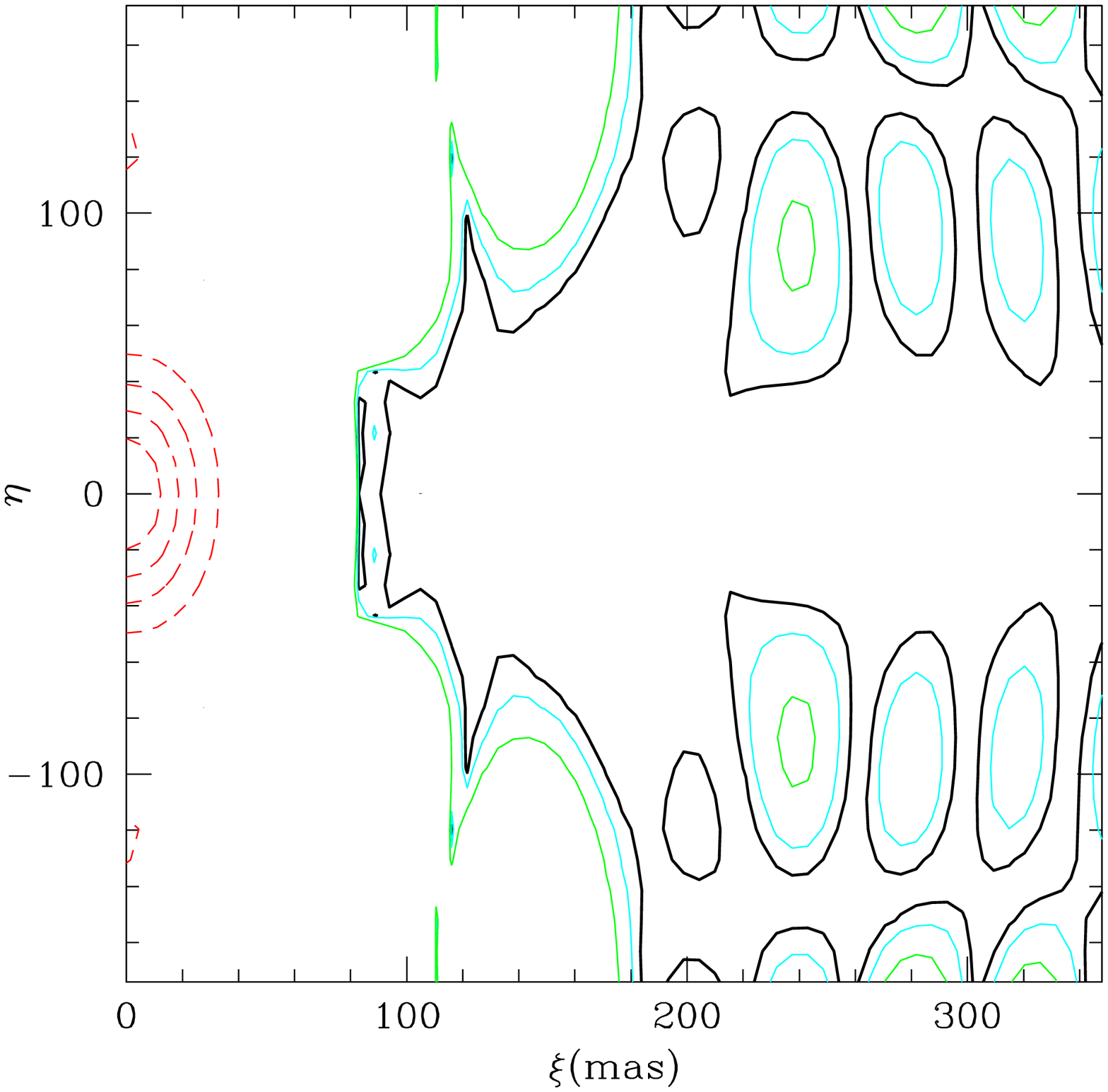}}}
\caption{The intensity response of the optimized pupil.  The dotted red lines
shows the shape of the central image.  The contours are  0.2, 0.4, 0.6
and 0.8 of the peak intensity.  The heavy black line encloses  the region
with intensity of less then $10^{-7}$ of the peak intensity.  The light 
lines enclose regions with intensity less than $10^{-6.5}$
and $10^{-6}$ of the peak intensity.   Recall that adding the nuller
reduces these intensity by another factor of 10$^{-3}$.
This plot is calculated for
$\lambda = 1.1 $ microns, the longest target wavelength.  The design
has been optomized to minimize the light receieved into a 80 milliarcsecond
wide slit along the axis.}
\end{figure}

Figure 4 shows the optimized pupil.  Figure 5 shows the on-axis intensity
as a function of wavenumber for the 10 meter system observing a planet at
the seperation of 0.8. The dotted line in figure 5 shows the integrated
energy recieved into a slit of width 0.02 arcseconds in a system using the
split pupil design described in the previous section.Figure 6 shows the
full two-dimensional response of the pupil.

\section{Scattered Light Problem}

The TPF optical system will not be perfect.Variations in the mirror
surface will 
lead to phase and amplitude errors. Following \cite{Brown}, we
can estimate these effects using scalar wave theory. Because of these
errors, equation (1) is now modified
\begin{eqnarray}
I(\xi ,\eta ,k) &=&\int_{A}dxdy\exp \left[ ik(x\xi +y\eta )+ikG(x,y)\right]
T(x,y) \\
&=&\tilde{A}(\xi ,\eta )+k\tilde{G}(\xi ,\eta )+\delta \tilde{A}(\xi ,\eta )
\nonumber
\end{eqnarray}
where $T(x,y)$ is the transmission of the system and $G(x,y)$ describes
variations in the mirror surface As long as he amplitude and phase errors
are signficantly less than unity, we can approximate the effect of phase
errors and amplitude errors as 
\begin{eqnarray}
\tilde{G}(\xi ,\eta ) &=&\int_{{}}dxdy\exp \left[ ik(x\xi +y\eta )\right]
G(x,y)  \nonumber \\
\delta \tilde{A}(\xi ,\eta ) &=&\int_{{}}dxdy\exp \left[ ik(x\xi +y\eta )%
\right] \left[ T(x,y)-\bar{T}\right] 
\end{eqnarray}
where .$\bar{T}$ is the average transmission. \ $\tilde{G}$ and $\delta 
\tilde{A}$ are complex functions. \ It is convienent for the next subsection
to split them into even and odd components.

While the engineering goal will be to minimize the errors in the mirror
surface, the tolerances are so severe that on-orbit corrections will be
essential. There are two complementary approaches to correcting the
amplitude and phase errors:

\begin{itemize}
\item  by slightly varying the shape of the aperture, we can cancel the
amplitude errors along $\eta =0$ and can also cancel $\partial \lbrack
\delta \tilde{A}(\xi ,0)]/\partial \eta $ 

\item  a deformable mirror\cite{Malbet} can correct phase
errors. Since the planned observations are primarily along a single axis,
the deformable mirror does not need to correct errors over the full surface,
but only along one dimension.
\end{itemize}
A detailed understanding of the mirror surface is essential for
characterizing mirror imperfections. In the next section, we show how we
can use observations with and without the nuller to characterize mirror
errors

\subsection{Using the nuller and spectrograph to measure errors}

Here, we envision an optical system consisting of the optimized pupil, an
X-shaped stop that blocks most of the scattered light, and a mirror that can
send the residual star pattern either through a nuller or directly into a
spectrograph.  Note that because of the stop, very little light from the
star enters the nuller, so that any optical errors in the mirrors in the
nuller will not contribute significantly to the scattered light problem.

When the beam goes through the nuller,  the spectrograph will measure 
\begin{eqnarray}
U_{nuller}(k,\xi ) &=&\left| \frac{1}{2}\left[ 1+\left( 2\pi k\Delta \right)
^{2}\right] I(\xi ,0)-\frac{1}{2}I(-\xi ,0)\right| ^{2}  \nonumber \\
&=&\left( k\tilde{G}_{odd}(\xi ,0)+\delta \tilde{A}_{odd}(\xi ,0)\right) ^{2}
 \nonumber \\
&&+2\left( 2\pi k\Delta \right) ^{2}\left( k\tilde{G}_{odd}(\xi ,0)+\delta 
\tilde{A}_{odd}(\xi ,0)\right) \left( \tilde{A}_{even}(\xi ,0)+k\tilde{G}%
_{even}(\xi ,0)+\delta \tilde{A}_{even}(\xi ,0)\right)  \nonumber \\ 
&&+O(\Delta )^{4}
\label{null}
\end{eqnarray}
On the other hand, when the signal does not pass through the nuller, the
detected signal will pick up both
the odd and even terms.

Planets will also produce a response in the system. Unlike
amplitude and phase errors, the effect of a planet is to produce a signal
that is centered at a constant value of $v$ rather than a constant value of $%
k\xi .$When we combine all the effects, the number of photons detected by
the optical system without the rooftop nuller is:
\begin{eqnarray}
U_{nonuller}(k,\xi ) &=&\left| I(\xi ,0)\right| ^{2}  \nonumber \\
&=&\left( \tilde{A}_{even}(\xi ,0)+k\tilde{G}_{even}(\xi ,0)+\delta \tilde{A}%
_{even}(\xi ,0)\right) ^{2}  \label{nonull}
\end{eqnarray}

If we ignore terms of order $\left( 2\pi k\Delta \right) ^{2},$ then we can
immediately compute the amplitude and phase errors from equations (\ref{null}%
) and (\ref{nonull}).

\section{Detectability of Earthlike Planets}
\begin{figure} [h]
\centerline{\scalebox{.4}{\includegraphics{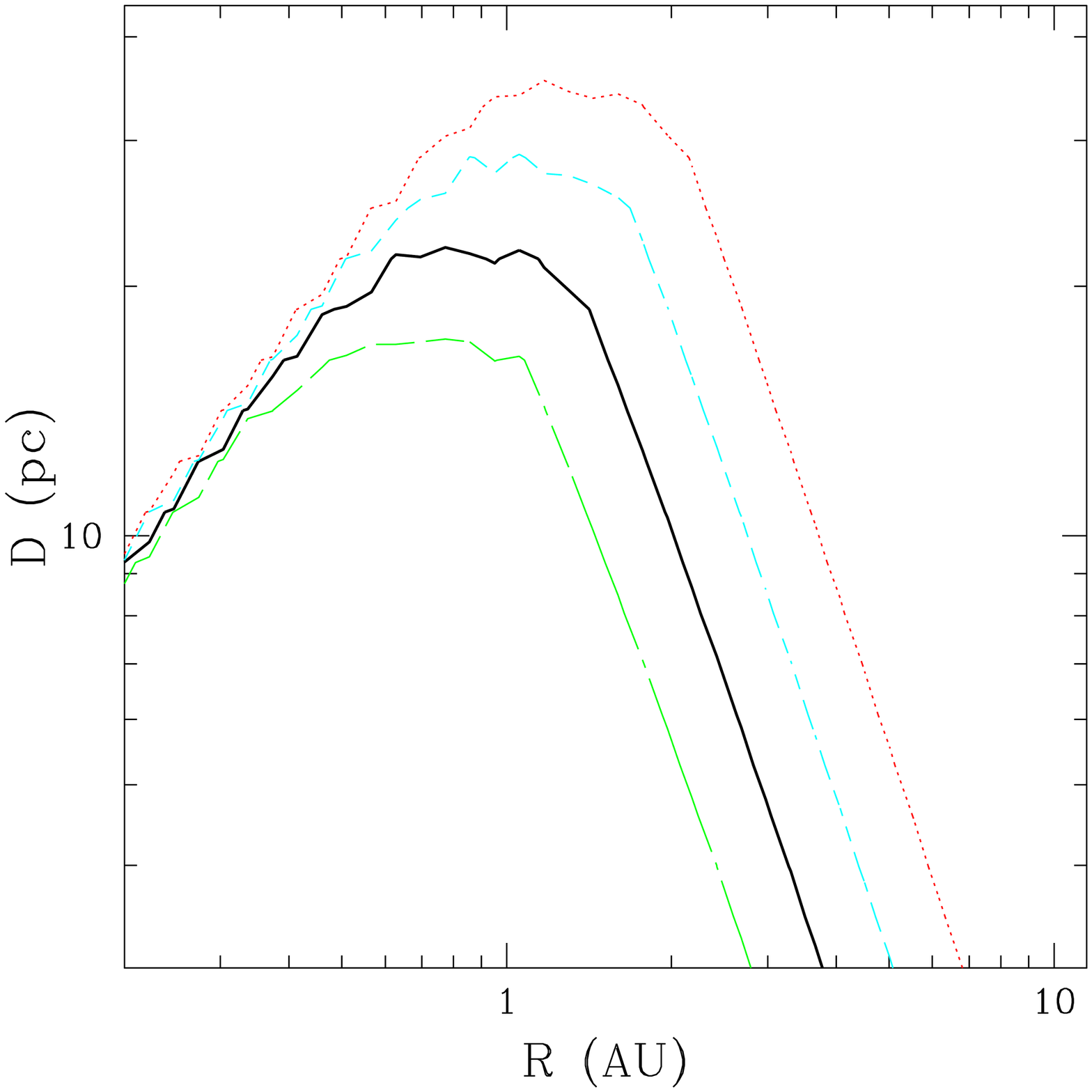}}}
\caption{This figure shows the time needed to detect a planet at the
5$\sigma$ level as a function of the distance to the host star
and the distance from the host star to the target planet.  The estimate
includes the time needed to rotate the telescope to scan the solar system.
The contours are for 3 Ks, 10 Ks, 30 Ks and 100 Ks integrations.
This right hand panel is for a solar type star.}
\end{figure}
\begin{figure} [h]
\centerline{\scalebox{.4}{\includegraphics{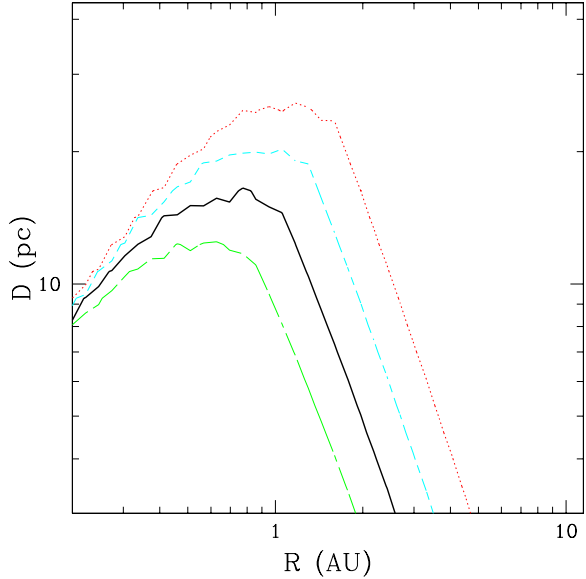}}}
\caption{This figure shows the time needed to detect a planet at the
5$\sigma$ level as a function of the distance to the host star
and the distance from the host star to the target planet.  The estimate
includes the time needed to rotate the telescope to scan the solar system.
The contours are for 3 Ks, 10 Ks, 30 Ks and 100 Ks integrations.
This panel is for a K2 host star.}
\end{figure}

In this concluding section, I  estimate the detectability of Earthlike
planets around G and K stars based on a system built from two 3 x 10 meter
mirrors and the system described in this note. In the estimate, I have
assumed that the deformable mirrors have been able to correct mirrors at the
level of $\lambda /10^{4},$ so that the amplitude of scattered light is
comparable to the amplitude of the diffracted light (10$^{-10}$ peak value).
With this level of supppresion, a planet 1 AU from the host star is as
bright as the background light (Q=1 in the language of \cite{Brown}). 
Since the brightness of the planet drops \ as the distance from the host
star increases, the Q value drops for planets further out in the system.

In estimating the signal-to-noise, I have assumed that the Earth-mass planet
was at optimal phase and an albedo of 0.5 regardless of frequency for the
planet's surface. I assumed that the planet was monitored in U, B,V, R and
I and combined the 5 bands to maximize the signal/noise. I assumed a 10\%
efficency for the full system at all wavelengths. For planets inside of
0.07 arcseconds, the shorter wavelengths are more important as diffracted
light renders the longer wavelengths unusable. While the star is dimmer at
short wavelengths, planets near the host star are brighter so that the two
factors nearly cancel. Figure 7 shows estimates of time needed to detect
the planet around a G2 and K2 star as a function of stellar distance and
seperation. This figure shows that the proposed system appears to be a
very effective planet finder, all of the G and K stars within 20 parsecs
could easily be surveyed for Earthlike planets during the mission lifetime.

The proposed system should also be able to characterize the atmospheres of
any detected planet. For an Earth twin around a star at a distance of 10
parsecs, the system should be capable of studying the atmosphere from 0.3 -
1.3 microns. This window contains a number
of very important biotracers (O$_{2},$O$_{3}$ and H$_{2}$O)\cite{Traub}.

A large optical telescope is a viable alternative to the traditional
mid-infrared approach to planet detection. Building a large space
telescope will be a challenging task; however, the potential scientific
rewards are enormous in fields ranging from cosmology to planetary science. 

\section{Acknowledgements}

This adventure into optical design would not have been possible without the
advice of my colleagues. I have benefited from discussions with Jim Gunn,
Norm Jarosik, Jeremy Kasdin, Michael Littman, Richard Miles, Sarah Seager
and Ed Turner at Princeton.  My fellows members of the Ball TPF team, in
particular,John Bally, Torsten Boecker, Bob Brown, Chris Burrows, Christ
Ftaclas, Steve Kilston, Charley Noecker, and Wes Traub, have provided
insightful criticism and commments and have deepened my understanding of
optics.

%% Figures and tables after References

\end{document}